\documentclass[preprint2]{aastex}
\newcommand{\myemail}{terada@phy.saitama-u.ac.jp}
\slugcomment{submitted 15 May 2010; accepted 4 Aug 2010}

\shorttitle{{\it Suzaku} Observation of AM Her in a very low state}
\shortauthors{Terada et al.}

\begin{document}


\title{X-ray observation of AM Herculis 
in a very low state with {\it Suzaku}}

\author{
Y. Terada\altaffilmark{1},
M. Ishida\altaffilmark{2,3}, 
A. Bamba\altaffilmark{2,4},
K. Mukai\altaffilmark{5},
T. Hayashi\altaffilmark{2,3}, and
A. Harayama\altaffilmark{1}
}
\email{\myemail}
\altaffiltext{1}{Department of Physics, Science, Saitama University, 
255 Simo-Ohkubo, Sakura-ku, Saitama city, Saitama 338-8570, Japan}
\altaffiltext{2}{Department of High Energy Astrophysics,
Institute of Space and Astronautical Science (ISAS),
Japan Aerospace Exploration Agency (JAXA),
3-1-1 Yoshinodai, Sagamihara, Kanagawa 229-8510, Japan}
\altaffiltext{3}{Science of Physics, Tokyo Metroporitan University,
1-1 Minami-Osawa, Hachioji-si, Tokyo, 192-0397, Japan}
\altaffiltext{4}{School of Cosmic Physics, 
Dublin Institute for Advanced Studies,
31 Fitzwilliam Place, Dublin 2,Ireland}
\altaffiltext{5}{Exploration of the Universe Division, Code 660,
NASA/GSFC, Greenbelt, MD 20771, USA}

\begin{abstract}
The X-ray observation of AM Her in a very low state was 
performed with {\it Suzaku} in October 2008.
One flare event with a time scale of $\sim$ 3700 sec
was detected at the X-ray luminosity of 
$6.0 \times 10^{29} {\rm ~erg ~sec}^{-1}$ in the 0.5 -- 10 keV band
assuming at a distance of 91 pc. The X-ray spectrum is represented by 
a thermal plasma emission model 
with a temperature of $8.67_{-1.14}^{+1.31}$ keV.
During the quiescence out of the flare interval, 
{\it Suzaku} also detected significant X-rays at a luminosity of
$1.7 \times 10^{29} {\rm ~erg ~sec}^{-1}$ in the 0.5 -- 10 keV band,
showing a clear spin modulation at a period of 0.1289273(2) days
at BJD 2454771.581.
The X-ray spectra in the quiescence were represented by 
a MEKAL + Power Law (PL) model or a single CEMEKL model,
which are also supported by phase-resolved analyses.
A correlation between the temperature and the volume emission 
measure was found together with 
historical X-ray measurements of AM Her in various states.
In order to account for a possible non-thermal emission from AM Her,
particle acceleration mechanisms in the AM Her system are also 
discussed, including a new proposal of a shock acceleration 
process on the top of the accretion column.

\end{abstract}

\keywords{novae, cataclysmic variables -- stars: individual (AM Herculis) 
-- plasmas -- acceleration of particles }

\section{Introduction}
\label{section:introduction}
Since white dwarfs (WDs) exist everywhere in our Galaxy
and have so large number density that one third of all the 
astronomical galactic objects are WDs \citep{allen73},
they have some possibilities to make important contributions
to unresolved long-standing mysteries in the astrophysics, 
such as the origin of Cosmic-ray particles, 
Galactic ridge emissions, and so on. 
Therefore, it is important to understand 
the nature of WDs in detail.
Magnetic cataclysmic variables (MCVs) are binary systems
which consist of late type stars and WDs with strong magnetic
field strength of over $10^5$ G. 
Cold gases from a companion star via a Roche-Lobe overflow 
accrete onto the magnetic pole(s) of the WD, and
release their gravitational potentials 
at the standing shock on the WD surface.
The temperature of the shock heated plasma, 
called an accretion column, reaches about $10^8$ K, 
and the plasma emits hard X-rays via bremsstrahlung process
\cite[see the review by][]{mcv_review}.
Polars are sub-class of MCVs, 
which have so strong magnetic field strength of $>10^7$ G
that spin periods of WDs are synchronized to the orbital periods.

In this paper, we report on the X-ray observation of
the proto-type star of polars, \object{AM Her}, 
with {\it Suzaku} \citep{suzaku07}.
The distance to the object $d$, the magnetic field strength 
on the surface of the WD, $B_{\rm WD}$, 
the mass of the WD, $M_{\rm WD}$, its radius, $R_{\rm WD}$, and 
the spin period, $P_{\rm spin}$, are measured as
$d=91$ pc \citep{gansicke95}, 
$B_{\rm WD} = $ 13--30 MG 
\citep{schmidt81, schmidt83, wichramasinghe84, wichramasinghe85},
$M_{\rm WD}$ = 0.39--1.2 $M_\odot$ 
\citep{young81, mukai87, bailey88, cropper98}, 
and 
$P_{\rm spin} = 3.09$ hr \citep{stockman77}, respectively, 
where $M_\odot$ and $R_\odot$ are 
the mass and radius of the Sun, respectively.
Due to changes of an accretion rate, $\dot{M}$, from the companion star,
the optical flux of AM Her shows two states;
the high state with V magnitude of 13.5 mag and 
the low state with 15.5 mag.
The previous X-ray observations in high states are performed
by {\it HEAO-1} \citep{rothschild81},
{\it EXOSAT} \citep{osborne86},
{\it Ginga} \citep{ishida91, beardmore95, cropper98},
and {\it ASCA} \citep{ishida97, terada04}.
They all report the detail measurements of the thermal emission
from the accretion column of AM Her.
On the other hand, X-ray observations in the low state are reported
only by {\it BeppoSAX} \citep{matt00}, 
and those in an intermediate state are covered 
by {\it Chandra} \citep{girish07}.

X-ray studies of thermal emissions in a low state can help us
to understand the physics of the accretion flow
by exploring one of the important physical parameters, 
$\dot{M}$, in a wide range.
In addition, in a very low state, 
it should be easier to search for unknown faint emissions 
behind the thermal radiation.
One of possible new-type components is a non-thermal emission as suggested 
by \citet{terada08b} from {\it Suzaku} observations of 
an intermediate polar AE Aqr, which belongs to 
a sub-class of MCVs having weaker magnetic field strength than polars.
Although the observation was actually not in a low state,
the high-sensitive wide-band spectroscopy and phase-resolved analyses
enabled them to find spiky pulsations like neutron star (NS) pulsars 
in the spin profiles with two independent instruments 
on-board {\it Suzaku}. 
They claimed that these spikes may have non-thermal origin
from detailed analyses of X-ray spectra. 
Thus, the object is now called as a ``NS pulsar equivalent of WD''
\citep{aeaqr_acc} or just a ``WD pulsar''.
X-ray deep observations in very low states could provide 
such a discovery; a second ``WD pulsar'', 
which is important in studies of searches for 
a new cosmic-ray acceleration system. 
In this paper, we describe the X-ray observation of AM Her 
in a low state in section \ref{section:obs},
summarize the results of {\it Suzaku} analyses 
during the flare and the quiescence in section \ref{section:ana}, 
and discuss in section \ref{section:discussion}. 

\begin{deluxetable}{rrrrrc}
\tablecolumns{4}
\tablewidth{0pc}
\tablecaption{Source list in the background region.}
\tablehead{
\colhead{OBSID} & \colhead{RA} & \colhead{DEC} &
\colhead{Flux$^\dagger$} & \colhead{index$^\ddagger$} &
\colhead{Identification}}
\startdata
403007010 & 18:17:10.379 & +49:47:06.95 
          & $4.80_{-1.13}^{+1.16}$ & $1.33_{-0.31}^{+0.34}$ 
          & 1WGA~J1817.1+4947\\
403007010 & 18:16:34.923 & +49:48:32.84 
          & $1.90_{-0.49}^{+0.49}$ & $2.57_{-0.65}^{+0.67}$ 
          & \object{SUZAKU~J1816.6+4948.5}$^\S$\\
403008010 & 18:48:04.720 & +47:57:34.24   
          & $6.73_{-1.13}^{+1.13}$ & $2.07_{-0.32}^{+0.31}$ 
          & \object{SUZAKU~J1848.0+4757.6}$^\S$ \\
403008010 & 18:48:17.747 & +47:52:25.25 
          & $2.41_{-2.40}^{+1.27}$ & $2.44_{-1.26}^{+1.42}$ 
          & \object{SUZAKU~J1848.3+4752.4}$^\S$ \\
\enddata
\label{tbl:background_src_pos}
{$\dagger$ X-ray flux in 0.5 -- 10 keV band, $10^{-14}$ erg sec$^{-1}$ cm$^2$.}\\
{$\ddagger$ Photon index of the object.}\\
{$^\S$ Newly discovered with {\it Suzaku} in this work. }\\
\end{deluxetable}

\section{Observation and Data Reduction}
\label{section:obs}
\subsection{{\it Suzaku} observation of AM Her}
\label{section:obs_obs}
{\it Suzaku} is the fifth series of the Japanese X-ray satellite,
carrying two X-ray instruments; 
the X-ray Imaging Spectrometer \citep[XIS;][]{xis2007}
and the Hard X-ray Detector \citep[HXD;][]{hxd2007a}.
It has good capabilities to provide very low-background observations 
with high sensitivities in the 0.2 -- 600 keV bandpass. 
Thus, it is suitable to search for a hidden emission
from a faint object with a wide-band spectroscopy.
In addition, high timing capability with the time resolution of 
8 sec and 64 $\mu$sec for the XIS and the HXD, respectively \citep{terada08a}, 
can help us to follow the fast variability of the object.

We observed AM Her with {\it Suzaku} from
2008 October 29 20:22~UT to November 1 07:18~UT
on the HXD-nominal position (OBSID=403007010; hereafter AMHER).
According to optical reports by
the American Association of Variable Star Observers (AAVSO)
\footnote{http://www.aavso.org}, 
this observation date corresponds to the low state of the object
lasting from September 2008 to June 2009.
In addition to the main observation tagged as AMHER,
we also performed an off-axis observation near the object 
to have the best use of the {\it Suzaku} sensitivities
from 2008 November 1 07:18~UT to November 2 08:15~UT
(OBSID = 403008010; hereafter AMHER\_OFF\_AXIS).
The off-axis observation was carried out towards 
the position of (RA, DEC)=(282.08, 47.99),
where no hard X-ray sources are contaminated
in the field of view of the HXD PIN and GSO.
On both occasions, the XIS was operated in the normal clocking mode
without window/burst options but with the Space Charge-Injection (SCI) 
function \citep{xis_sci08}.
The HXD was operated in the nominal mode; half of 64 PIN diodes
were operated with a voltage of 400 V and the others with 500 V. 

\subsection{Data Reduction}
\label{section:obs_reduction}
We used the data sets by the standard {\it Suzaku} pipe-line processing
version 2.2.11.22, with the calibration version (CALDBVER) of 
hxd20081009, xis20081009, xrt20080709, and xrs20060410.
We used the ftools in HEADAS 6.6 with XSPEC version 11.3.2ag.

The source was detected with the XIS 
at the average count rate of 
0.03 -- 0.04 cnt sec$^{-1}$ per sensor in the 0.5 -- 10 keV band.
Cleaned events of the XIS data were obtained with the standard criteria 
of the pipe-line process. The exposures of AMHER and AMHER\_OFF\_AXIS
observations were 108.5 ksec and 44.3 ksec, respectively.
On-source events of the XIS were then obtained 
by accumulating within 6 mm (4.3') radius 
from the image centroid of the object in the AMHER dataset.
Several faint sources (including one known object)
were detected both in AMHER and AMHER\_OFF\_AXIS datasets
with the XIS count rate of less than about $0.005$ cnt sec$^{-1}$,
which is comparable to the count rate of the background and 
corresponds to the X-ray flux of $0.7 \times 10^{-13}$ erg sec$^{-1}$ cm$^2$ 
as listed in table \ref{tbl:background_src_pos}.
Thus, the background events were accumulated by source free regions
of AMHER and AMHER\_OFF\_AXIS,
where $>6$ mm outer than these faint sources or AM Her.
We summed up these two background spectra, 
because these spectral shapes and the flux of 
the two blank sky datasets 
were consistent within 5 \% in the 0.5 -- 10 keV range.

In the data reduction of the HXD, 
we first reprocessed the unscreened events by the tool 'hxdpi'
and 'hxdgrade' \citep{terada05} using the CALDB files 
(including the GSO gain-history file named 
ae\_hxd\_gsoght\_20090131.fits to correct the PMT gains of GSO), 
and then obtained cleaned events of PIN and GSO 
with the standard criteria of the process. 
The exposures of the HXD of AMHER and AMHER\_OFF\_AXIS were
94.4 ksec and 40.3 ksec, respectively.
The non X-ray background (NXB) events of PIN and GSO are estimated
and provided by the HXD team with the methods by Fukazawa et al.\ (2009).
We used NXB events of both PIN and GSO with 
METHOD=`LCFITDT (bgd\_d)' and version of METHODV='2.0ver0804'.
The systematic errors of the NXB can be checked by comparison 
between the actual data and the NXB data during the night earth observations.
In AMHER and AMHER\_OFF\_AXIS observations, 
from night earth spectra with exposures of 38 ksec and 59 ksec 
for PIN and GSO, respectively,
the reproducibilities of PIN and GSO NXBs were 
about 4 and 1 \%, respectively.
Within these systematic errors, the events detected with PIN were
consistent with the Cosmic X-ray background (CXB) level,
and those with GSO were consistent with being no signals.

\section{Analyses and Results}
\label{section:ana}
\subsection{X-ray Light Curves}
\label{section:ana_lc}
\begin{figure}[hbt]
\centerline{\includegraphics[angle=0,scale=.15]{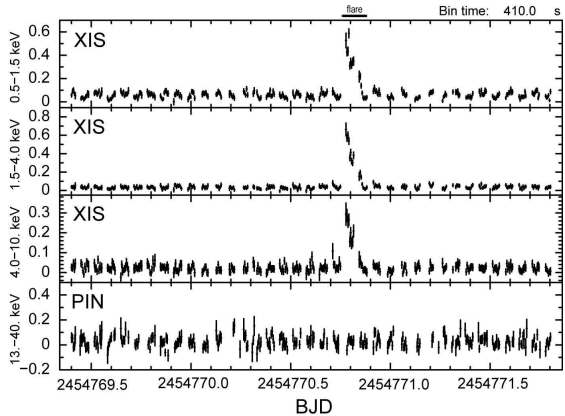}}
\caption{Background subtracted light curves taken with the {\it Suzaku} 
XIS and HXD in 0.5 -- 1.5, 1.5 -- 4.0, 4.0 -- 10.0, 13 -- 40 keV band, 
from top to bottom, respectively.}
\label{fig:light_curve}
\end{figure}

Fig.\ \ref{fig:light_curve} shows background-subtracted light curves 
of AM Her, obtained with the XIS and the HXD. 
The arrival times in this figure are barycentric 
corrected with the tool named 'aebarycen' \citep{terada08a}.
In the XIS data, a flare phenomenon starting from BJD 2454770.78 was
observed with a duration of about 7 ksec. 
There was not any brightening in the background data of the XIS,
and so the flare should arise from AM Her itself.
On the other hand, the PIN data showed no significant flare
synchronized to the XIS one.
Here, we defined the duration between BJD 2454770.777416 -- 2454770.863308
as an epoch of the flare, and the others as the quiescence.
The effective exposures of the XIS during the flare and the quiescence 
were 5.0 ksec and 103 ksec, respectively.

\begin{figure}[htb]
\centerline{\includegraphics[angle=0,scale=.20]{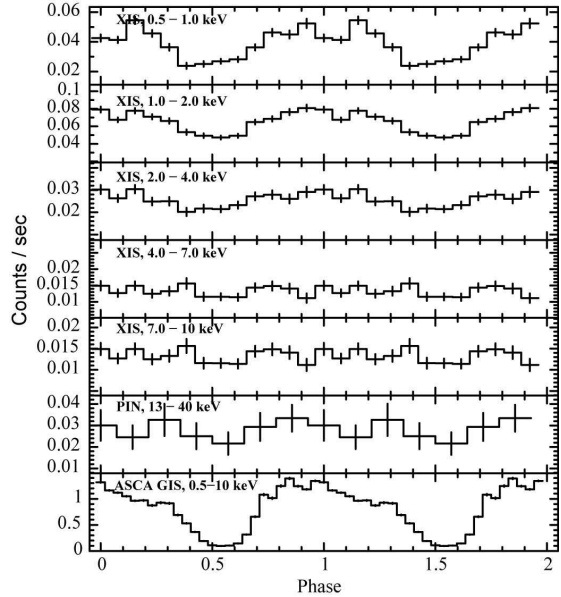}}
\caption{Light curves folded at the spin period of $P_{\rm XIS} = $ 0.1289273
days shown in 0.5 -- 1.0, 1.0 -- 2.0, 2.0 -- 4.0, 
4.0 -- 7.0, 7.0 -- 10.0, and 13 -- 40 keV band 
from the top to bottom panels, respectively. The phase 0.0 corresponds to 
BJD 2454771.581. The upper six panels are taken with {\it Suzaku}, 
and the lowest panel shows the same plot but with {\it ASCA} GIS in 1993
\citep{terada04}.}
\label{fig:efold}
\end{figure}

We search for the spin period from the XIS data in the 0.5 -- 10 keV band
during the quiescence, and obtained the period 
$P_{\rm XIS} = 0.1289273(2) $ days at the epoch BJD 2454771.581.
The value is consistent with the $P_{\rm spin}$ by 
previous observations \citep{stockman77, kafka05}; 
i.e., $P_{\rm XIS} = P_{\rm spin}$.
Here, the epoch was simply determined by the peak time
of the pulse profile fitted with the sine function.
Fig.\ \ref{fig:efold} shows the energy-resolved light curves
folded at $P_{\rm XIS}$. Clear spin modulations are observed 
especially in the soft energy band below 4 keV.
They have similar pulse shapes to those taken with 
{\it ASCA} \citep{asca1994} during the high state,
although significant energy dependence of the spin amplitudes is 
detected, unlike those in high states \citep{ishida97}.

\subsection{X-ray Properties during the Flare}
\label{section:ana_flare}
\begin{deluxetable}{lcccccc}
\tablecolumns{2}
\tablewidth{0pc}
\tablecaption{Best fit model parameters in the flare epoch of AM Her.}
\tablehead{
\colhead{Model} &
\colhead{$n_{\rm H}$} 
        & \colhead{$kT$} & \colhead{$\alpha$ $^\ddagger$} & 
\colhead{abundance} &
\colhead{Flux$^\dagger$} & \colhead{$\chi^2_\nu$ (d.o.f)} \\
\colhead{} & 
\colhead{$10^{20}$ cm$^{-2}$} & \colhead{keV} & \colhead{} & 
\colhead{solar}&
\colhead{$10^{-12}$ erg cm$^{-2}$ s$^{-1}$} & \colhead{}}
\startdata
phabs * MEKAL 
        & $< 1.3$ 
        & 8.67$_{-1.14}^{+1.31}$ & \nodata   
        & 0.76$_{-0.26}^{+0.28}$ 
        & 5.96$_{-0.26}^{+0.11}$ & 1.00 (205) \\
phabs * CEMEKL
        & $< 4.0$ 
        & 17.6$_{-6.6}^{+13.9}$ & 1.64$_{-0.65}^{+0.70}$
        & 0.90$_{-0.35}^{+0.40}$ 
        & 6.05$_{-0.38}^{+0.02}$ & 0.99 (203) \\
\enddata
\\
$\dagger$ X-ray flux in 0.5 -- 10 keV band.\\
$\ddagger$ The power of $DEM$, as presented by 
$DEM \propto (T/T_{\rm max})^{\alpha-1} d(T)$ 
in the CEMEKL model \citep{done97}.
\label{tbl:spec_flare}
\end{deluxetable}

\begin{figure}[h]
\centerline{\includegraphics[angle=0,scale=.15]{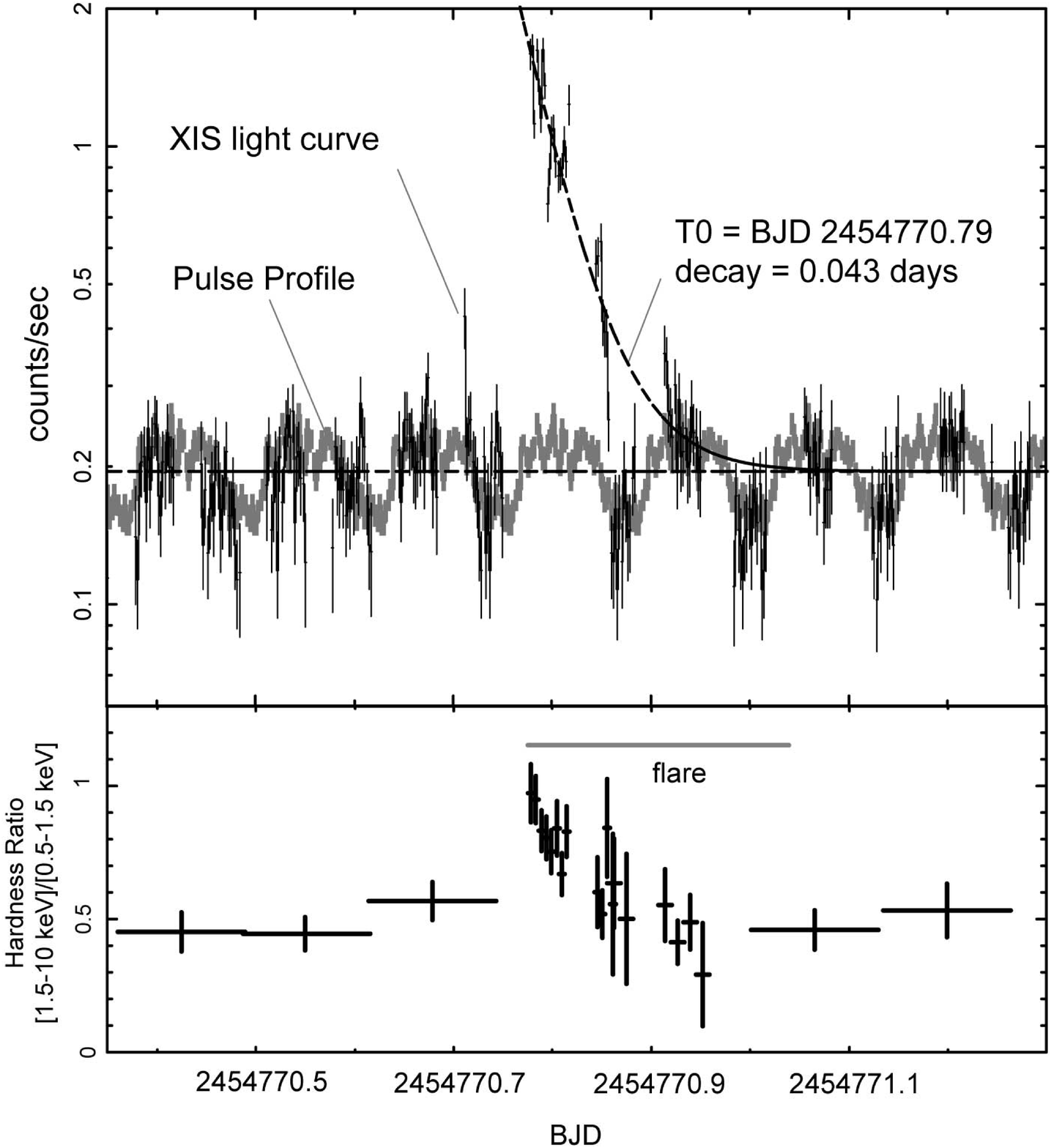}}
\caption{Top panel shows the light curves of the XIS in the 0.5 -- 10 keV 
band near the flare event shown in black crosses, 
and pulse profiles of the XIS in quiescence folded at the spin period of 
$P_{\rm XIS} = $ 0.1289273 days are also shown in gray.
The dashed lines show the average count of the quiescence and 
the best-fit exponential curve (see the text). 
Bottom panel represents the hardness ratio of the XIS; 
counts in 1.5 -- 10.0 keV band divided by those in 0.5 -- 1.5 keV band.}
\label{fig:flare_lc}
\end{figure}

Fig.\ \ref{fig:flare_lc} shows the light curve of the XIS 
in the 0.5 -- 10 keV band around the flare. 
The folded 0.5 - 10 keV light-curve in the quiescence 
(like Fig.\ \ref{fig:efold}) are overlaid periodically for comparison.
If we subtract this spin variation,
X-ray counts around the phase bottom (BJD 2454770.86 -- 2454770.88)
became almost zero, but the others showed an exponential decay.
Then, we fitted the light curve without the phase-bottom epoch
by the exponential function of $\exp(({\rm T}- T_0) / \tau_{\rm flare})$,
where $T$, $T_0$, and $\tau_{\rm flare}$ are time,
starting time of the flare, and a decay constant of time, respectively. 
We obtained the time origin of $T_0 =$ BJD 2454770.76 $\pm$ 0.03 and 
the decay constant of $\tau_{\rm flare} = (4.3 \pm 0.2) \times 10^{-2}$ days.

\begin{figure}[htb]
\centerline{\includegraphics[angle=-90,scale=.35]{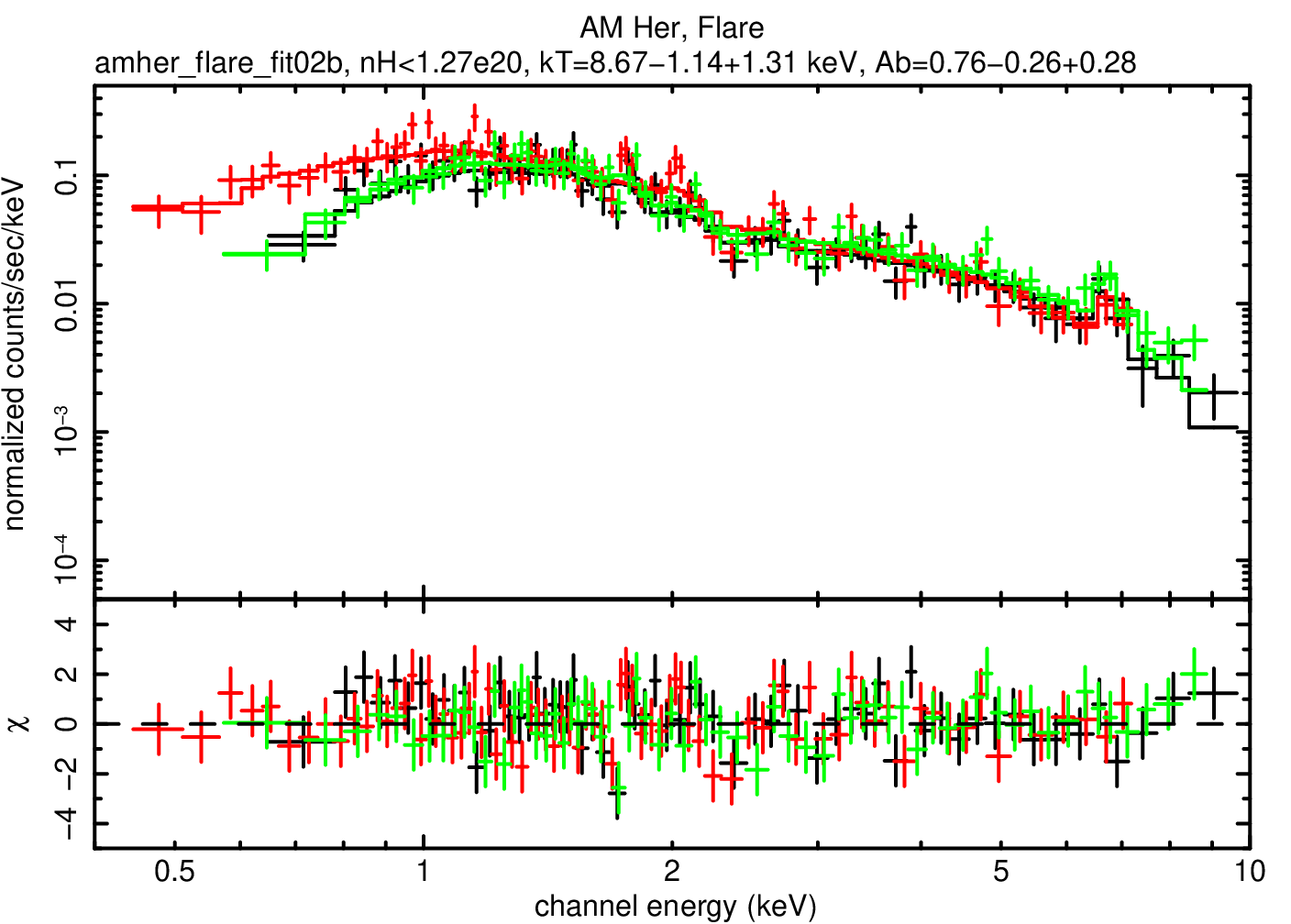}}
\caption{X-ray spectra with {\it Suzaku} during the flare of AM Her in the 
very low state. The black, red, and green crosses represents the data
of XIS-0, XIS-1, and XIS-3, respectively. Lines show the best fit model 
of MEKAL with parameters in table \ref{tbl:spec_flare}.}
\label{fig:spec_flare}
\end{figure}

The X-ray spectra during the flare are shown in Fig.\ \ref{fig:spec_flare}.
Since there were no significant detections by the PIN and GSO
within systematic errors of the NXB (section \ref{section:obs_reduction}), 
only XIS events were plotted in the figure.
First, we tried to reproduce the X-ray spectra by an emission model
from an optically-thin thermal plasma, which can be generated
at the post-shock region of the accretion gas-flow 
onto the WD via energy releases of their 
gravitational potential. The spectra obtained with {\it Suzaku} were
well reproduced by such a thin-thermal plasma model called MEKAL 
\citep{mewe85, liedahl95, spex}, as shown in table \ref{tbl:spec_flare}
and Fig.\ \ref{fig:spec_flare}.
The fitting to the XIS-0, 1, and 3 datasets were performed simultaneously,
and cross normalizations were consistent with the value of 1.0.
Although the temperature, $kT = 8.67_{-1.14}^{+1.31}$ keV, 
obtained during this very low state, was slightly lower than 
those in high states 
\citep[e.g., 13.5 keV or 11 keV by][respectively]{beardmore95,ishida97}
the metal abundance, $0.76_{-0.26}^{+0.28}$ solars, was consistent
with values in high states. Thus, the hot plasma would come from 
the accretion matters from the companion star.
The spectra requires no significant photo-electric absorption as shown in 
table \ref{tbl:spec_flare}; this fact is also self-consistent with 
the situation of a very low accretion rate.

You may notice that there remains a line structure around 2.0 keV 
in the residual plot of Fig.\ \ref{fig:spec_flare},
although they were insignificant.
If it was real, the features should come from a light element Si,
from which emission should be suppressed in such a high temperature 
as $kT = 8.67$ keV.
For a reference, we further tried an one-dimensional cooling-flow model 
as originally proposed by \citet{hoshi73} and \citet{aizu73},
and found that no more free parameters improve the fitting statistically.
The spectra were also well fitted by such a model as the CEMEKL model
\citep{done97, baskill05}, 
but there was no large improvement in the chi-squared value,
as shown in table \ref{tbl:spec_flare}.
In this fitting, the shock temperature, $T_{\rm max}$, were obtained
as $17.6_{-6.6}^{+13.9}$ keV, which is higher than 
the average temperature by the single MEKAL fitting,
because $T_{\rm max}$ represents the maximum temperature
just below the shock front in the accretion plasma.
The value itself was reasonable compared with the gravitational 
potential of the WD.
The power of the temperature gradient, $\alpha$, of 
the differential emission measure, $DEM$, was 
$\alpha = 1.64_{-0.65}^{+0.70}$, which was different from 
the theoretical expectation of 0.5 \citep{ishida94}
with a simple assumption of an one-dimensional stable 
cooling-flow by \citet{aizu73}

\begin{figure*}[htb]
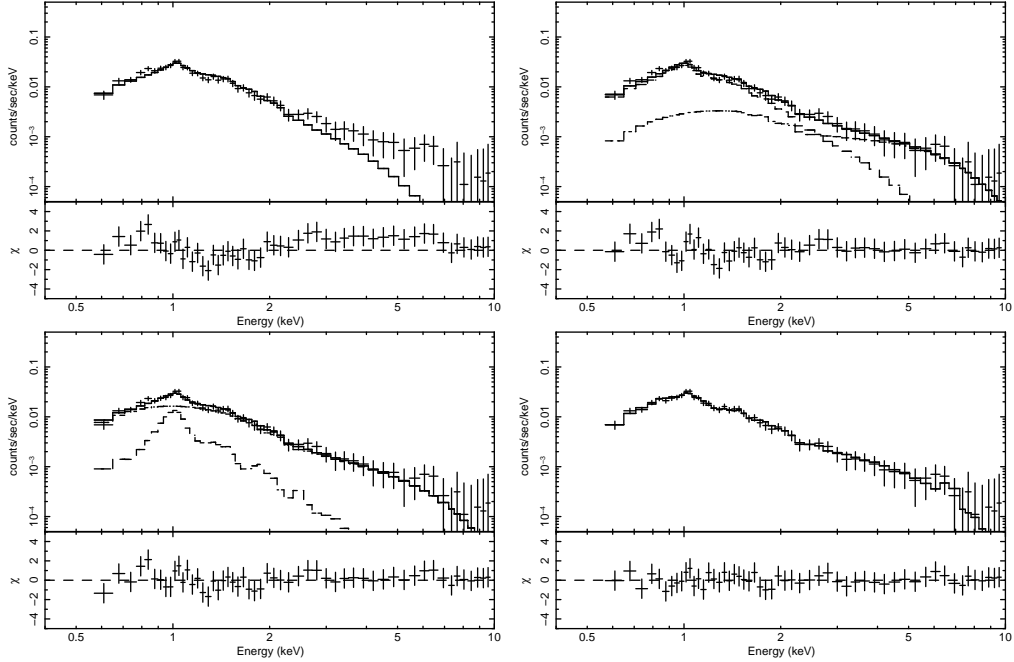

\centerline{
\includegraphics[angle=-90,scale=.28]{figure5a.eps}
\includegraphics[angle=-90,scale=.28]{figure5b.eps}
}
\centerline{
\includegraphics[angle=-90,scale=.28]{figure5c.eps}
\includegraphics[angle=-90,scale=.28]{figure5d.eps}
}
\caption{X-ray spectra with {\it Suzaku} during the quiescence of AM Her 
in the very low state. The crosses represents the XIS data summed of 
XIS-0, XIS-1, and XIS-3. Lines show the best fit models of single MEKAL, 
double MEKAL, MEKAL + power law, and CEMEKL, in top left, top right, 
bottom left and bottom right panels, respectively (see the text). 
Parameters of the models are listed in table \ref{tbl:spec_qui}.}
\label{fig:spec_qui}
\end{figure*}

\subsection{Average Spectral Analysis in the Quiescence}
\label{section:ana_qui_spec}
The X-ray spectra during the quiescence with {\it Suzaku} 
were shown in Fig.\ \ref{fig:spec_qui}. 
Cross normalizations between XIS-0, XIS-1, and XIS-3 were consistent with 1.0,
as already checked in section \ref{section:ana_flare},
and thus datasets of all the XIS chips were summed in the figure.
As implied in section \ref{section:ana_lc} and Fig.\ \ref{fig:efold}, 
the spin pulse profile during the quiescence shows the similar modulations
as in the high states, soft X-ray emissions should have thermal origin.
Therefore, we tried to represent the spectrum 
by a single MEKAL model, following the steps on the analyses during the flare
(section \ref{section:ana_flare}), 
but the fitting was not acceptable ($\chi^2_\nu = 1.30$)
as shown in table \ref{tbl:spec_qui}.
The metal abundance was quite low as 0.06$_{-0.02}^{+0.03}$ solar,
and so the best fit model has no physical meanings.

\begin{figure*}[htb]
\centerline{
\includegraphics[angle=0,scale=.15]{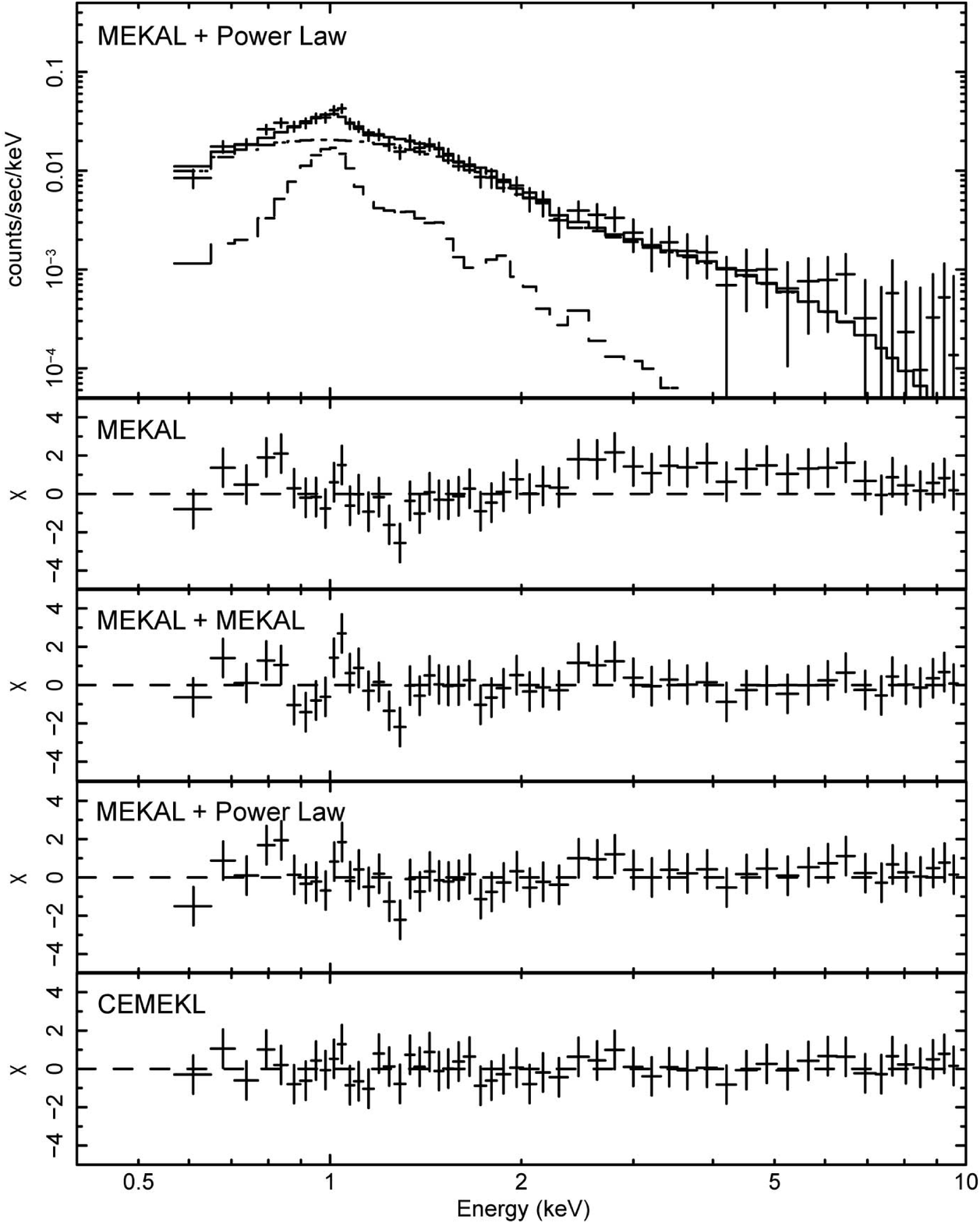}
\includegraphics[angle=0,scale=.15]{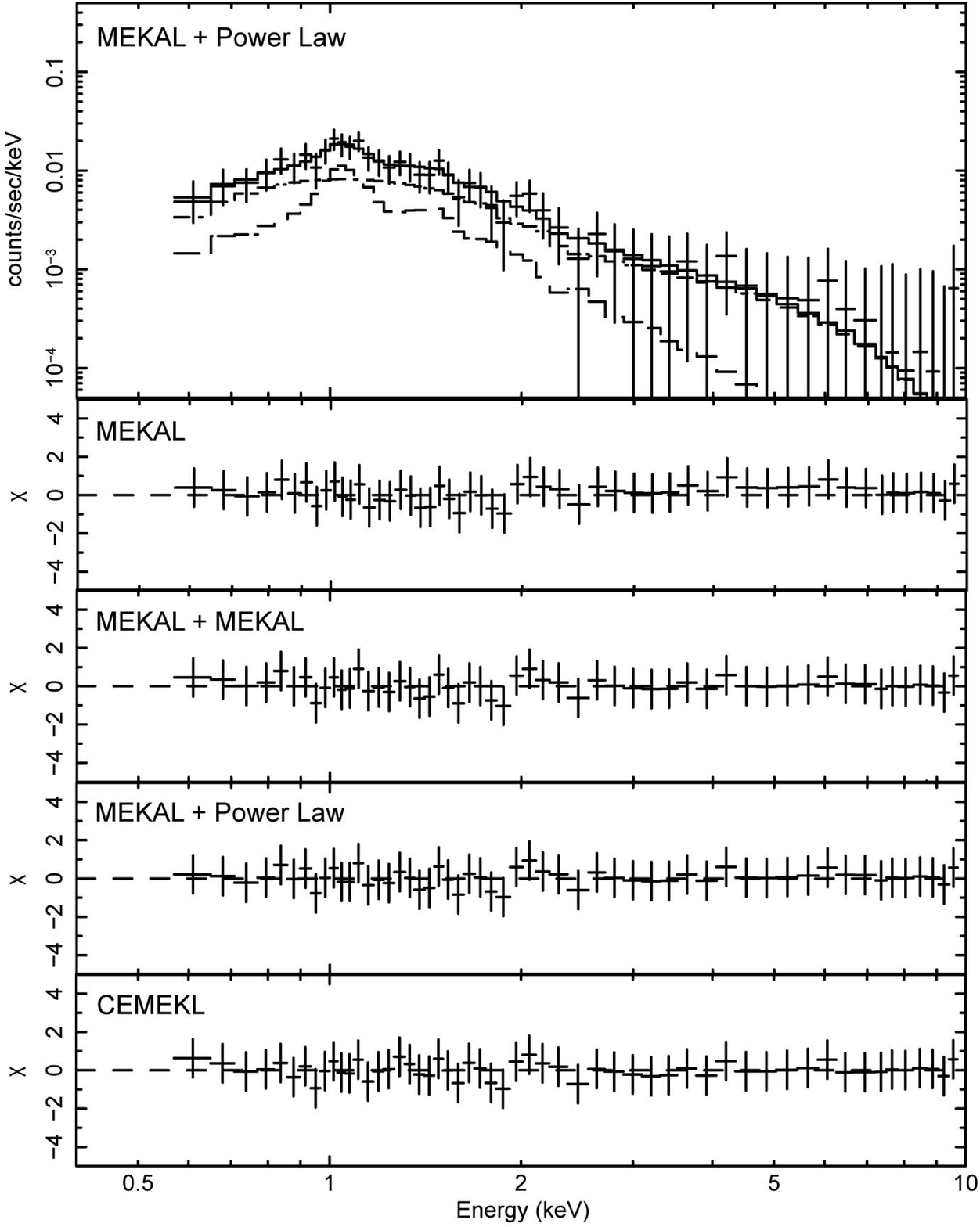}
}
\caption{Left and right panels show the X-ray spectra in the top and bottom
phases, respectively. Crosses are the XIS data. The top panel
in each figure shows the spectrum with the best fit model of MEKAL +
power law, and bottom panels are the residuals of data from 
models, MEKAL, MEKAL+MEKAL, MEKAL+Power Law, and CEMEKL, 
towards lower panels. Parameters of the models are summarized in 
table \ref{tbl:spec_qui}.}
\label{fig:spec_qui_resolved}
\end{figure*}

In the hard X-ray band, the spin pulse profiles (Fig.\ \ref{fig:efold})
showed shallower modulations than those in the softer energy band;
the pulse shape was consistent with being flat in above 10 keV range.
Thus, in the spectral analyses, we then added another component 
in the hard X-ray band, representing another thermal emission (case-1)
or non-thermal component (case-2).

In the first case that hard X-rays have a thermal origin (case-1),
we tried a double MEKAL model to represent the spectrum.
As a result, the fitting became acceptable ($\chi^2_\nu = 0.76$),
as shown in table \ref{tbl:spec_qui} and Fig.\ \ref{fig:spec_qui}.
However the temperature of the higher component was quite high as $>17.7$ keV
and the metal abundance was still too low as 0.06$_{-0.02}^{+0.02}$ 
solar, compared with the result during the flare 
(section \ref{section:ana_flare}) and/or 
reports by previous observations in high states 
\citep{ishida97, terada04, girish07}.
In the fitting, the two parameters of the abundance in both MEKAL
components were linked, but the situation did not change even when
we set these two parameters free from each other;
$1.03_{-0.06}^{+0.05}$ keV plasma with abundance of 
$0.06_{-0.01}^{+0.01}$ solar and
$>14.8$ keV plasma with $<0.85$ solar abundance ($\chi^2_\nu = 1.01$).

In the second case that hard X-rays have a non-thermal origin (case-2),
we tried MEKAL + PL model and obtained an acceptable result
($\chi^2_\nu = 0.61$), 
as shown in table \ref{tbl:spec_qui} and Fig.\ \ref{fig:spec_qui}.
The metal abundance of the MEKAL component, 0.80$_{-0.15}^{+0.40}$ solar,
became consistent with previous observations \citep{ishida97, terada04, 
girish07} and this work during the flare (section \ref{section:ana_flare}).
The photon index of the non-thermal component was 2.36$_{-0.14}^{+0.15}$.
The X-ray flux in the 0.5 -- 10 keV band of the non-thermal component
was $(1.5 \pm 0.1) \times 10^{-13}$ erg sec$^{-1}$ cm$^{-2}$,
which corresponds to the luminosity in the same energy band of
$(1.5 \pm 0.1) \times 10^{29} \left(\frac{d}{91 {\rm pc}}\right)^{2}$ 
erg sec$^{-1}$.
Note that a hump structure around 6 -- 7 keV in the residual panel
of Fig.\ \ref{fig:spec_qui} bottom left was statistically insignificant;
if we tentatively add a Gaussian model here to MEKAL + PL model, 
we obtained the central energy of the Gaussian at $6.3^{+2.1}_{-6.3}$ keV, 
where the improvement of $\chi^2_\nu$ was only 0.02
corresponding to the F-test value and the probability of 
$1.76$ and $0.18$, respectively.

As an application of case-1 fitting (double MEKAL model), 
we also tried the multi-color plasma model, CEMEKL, 
as already used in section \ref{section:ana_flare} (case-3), 
and got an acceptable result ($\chi^2_\nu = 0.32$)
as summarized in table \ref{tbl:spec_qui} and Fig.\ \ref{fig:spec_qui}.
The metal abundance and the $kT_{\rm max}$ of the best fit model
were reasonable as 0.48$_{-0.21}^{+0.22}$ solar and 
6.37$_{-1.37}^{+8.34}$ keV, respectively.
The power $\alpha$ of the Volume Emission Measure (VEM) was 
$\alpha = 0.58_{-0.18}^{+0.31}$,
which was consistent with the theoretical value of $\alpha = 0.5$ 
\citep{ishida94} in an ideal situation 
of the one-dimensional steady cooling flow.
In summary, the phase-averaged X-ray spectrum of AM Her in the quiescence
were reproduced not by single MEKAL model, 
but well by three kinds of models; 
(case-1) MEKAL + MEKAL, (case-2) MEKAL + PL, and (case-3) CEMEKL,
although case-1 requires an unrealistically low metal abundance.

\begin{deluxetable}{lccccccc}
\tablecolumns{2}
\tablewidth{0pc}
\tablecaption{Best fit model parameters during the quiescence of AM Her.}
\tablehead{
\colhead{Model} &
\colhead{$kT_1$} & \colhead{$kT_2$ or $kT_{\rm max}$} & 
\colhead{$\alpha^\dagger$ or $\Gamma ^\ddagger$} & 
\colhead{abundance} & \colhead{Flux$^\S$} & 
\colhead{$\chi^2_\nu$ (d.o.f)} \\
\colhead{} & 
\colhead{keV} & \colhead{keV} & \colhead{} & 
\colhead{solar}& \colhead{} & 
\colhead{}}
\startdata
\multicolumn{3}{l}{\bf Phase Average}& & & \\
MEKAL & 
1.23$_{-0.13}^{+0.09}$ & \nodata & \nodata & 
0.06$_{-0.02}^{+0.03}$ & 1.13$_{-0.11}^{+0.08}$ & 
1.30 (51) \\
MEKAL + MEKAL& 
1.02$_{-0.11}^{+0.07}$ & $>11.1$ & \nodata & 
0.06$\pm 0.02$ & 1.79$_{-0.30}^{+0.21}$ & 
0.76 (49) \\
MEKAL + PL& 
1.10$\pm 0.09$ & \nodata & 2.36$_{-0.14}^{+0.15}$  & 
0.80$_{-0.15}^{+0.40}$ & 1.73$_{-0.08}^{+1.27}$ & 
0.61 (49) \\
CEMEKL & 
  \nodata & 6.57$_{-2.57}^{+3.63}$ & $0.58_{-0.18}^{+0.31}$ & 
0.48$_{-0.21}^{+0.22}$ & 1.76$_{-0.54}^{+0.20}$ & 
0.32 (50) \\
\multicolumn{3}{l}{\bf Top Phase}& & & \\
MEKAL & 
1.07$_{-0.07}^{+0.06}$ & \nodata & \nodata & 
0.05$\pm 0.02$ & 1.62$_{-0.13}^{+0.14}$ & 
1.24 (51) \\
MEKAL + MEKAL& 
0.84$_{-0.08}^{+0.14}$ & $>6.2$ & \nodata & 
0.04$\pm 0.02$ & 2.36$_{-0.48}^{+0.14}$ & 
0.73 (49) \\
MEKAL + PL& 
1.04$_{-0.09}^{+0.07}$ & \nodata & 2.42$_{-0.18}^{+0.17}$  & 
0.80$_{-0.52}^{+0.56}$ & 2.13$_{-0.08}^{+1.85}$ & 
0.67 (49) \\
CEMEKL & 
  \nodata & 6.37$_{-1.37}^{+8.34}$ & $0.53_{-0.33}^{+0.19}$ & 
0.50$_{-0.24}^{+0.57}$ & 2.17$_{-0.62}^{+0.23}$ & 
0.37 (50) \\
\multicolumn{3}{l}{\bf Bottom Phase}& & & \\
MEKAL & 
1.68$_{-0.37}^{+0.60}$ & \nodata & \nodata & 
0.14$_{-0.10}^{+0.26}$ & 0.97$_{-0.22}^{+0.86}$ & 
0.24 (51) \\
MEKAL + MEKAL& 
1.37$_{-0.35}^{+0.59}$ & 79.9$_{-79.8}^{+0.1}$ & \nodata & 
0.11$_{-0.09}^{+2.74}$ & 1.27$_{-0.31}^{+0.45}$ & 
0.20 (49) \\
MEKAL + PL& 
1.43$_{-0.36}^{+3.07}$ & \nodata & 2.07$_{-1.65}^{+6.66}$  & 
0.38$_{-0.31}^{+0.49}$ & 1.22$_{-0.14}^{+0.26}$ & 
0.19 (49) \\
CEMEKL & 
  \nodata & 5.76$_{-2.38}^{+13.0}$ & $1.29_{-3.50}^{+0.73}$ & 
0.76$_{-0.67}^{+3.17}$ & 1.26$_{-0.56}^{+1.11}$ & 
0.17 (50) \\
\enddata
\\
$\dagger$ Valid for CEMEKL model, the power of $DEM$.\\
$\ddagger$ Valid for Power law model, the photon Index.\\
$\S$ The X-ray flux in 0.5 -- 10 keV band 
in $10^{-13}$ erg cm$^{-2}$ s$^{-1}$.\\
\label{tbl:spec_qui}
\end{deluxetable}

\begin{deluxetable}{lcccccc}
\tablecolumns{2}
\tablewidth{0pc}
\tablecaption{Best fit model parameters of the pulsation spectrum 
during the quiescence.}
\tablehead{
\colhead{Model} &
\colhead{$kT$} & \colhead{$\alpha ^\dagger$} & 
\colhead{$\Gamma ^\ddagger$} & 
\colhead{abundance} & \colhead{Flux$^\S$} & 
\colhead{$\chi^2_\nu$ (d.o.f)} \\
\colhead{} & 
\colhead{keV} & \colhead{} & \colhead{} & 
\colhead{solar}& \colhead{} & 
\colhead{}}
\startdata
PL& 
\nodata & \nodata & 
3.03$_{-0.39}^{+0.41}$  & \nodata & 1.00$_{-0.14}^{+0.21}$ & 
0.47 (52) \\
MEKAL & 
0.78$\pm 0.17$ & \nodata & \nodata &
0.05$_{-0.03}^{+0.08}$ & 0.73$_{-0.23}^{+0.16}$ & 
0.32 (51) \\
CEMEKL & 
10.8$_{-10.8}^{+89.2}$ & $0.12_{-0.12}^{+19.9}$ & \nodata &
0.84$_{-0.64}^{+9.16}$ & 1.00$_{-0.78}^{+5.47}$ & 
0.27 (50) \\
\enddata
\\
$\dagger$ Power of the $DEM$ in the CEMEKL model, $\alpha$. \\
$\ddagger$ Photon Index of the PL model.\\
$\S$ X-ray flux in the 0.5 -- 10 keV band in $10^{-13}$ erg cm$^{-2}$ s$^{-1}$.
\label{tbl:spec_qui_pulse}
\end{deluxetable}

\subsection{Phase resolved analyses in Quiescence}
\label{section:ana_qui_phaseresolved}
In order to restrict the emission model in the quiescence, 
by a combination of results of the timing analyses 
(section \ref{section:ana_lc})
and the spectral ones (section \ref{section:ana_qui_spec}),
we compared the best fit models of the phase-resolved spectra
with the spin pulse modulation in Fig.\ \ref{fig:efold}, quantitatively.

\begin{figure}[htb]
\centerline{
\includegraphics[angle=0,scale=.15]{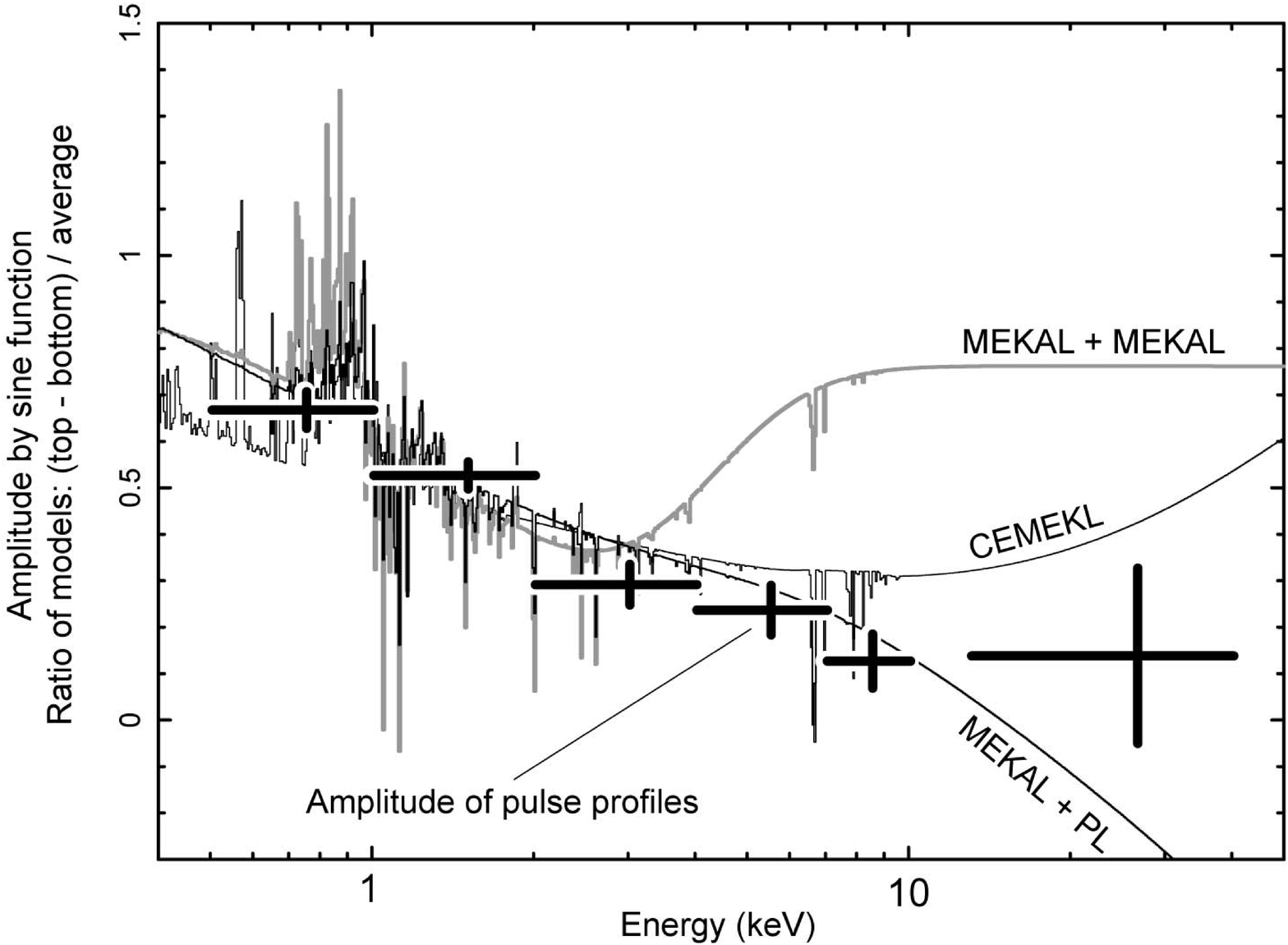}
}
\caption{Crosses represent the amplitudes of the energy-resolved 
pulse profiles in Fig.\ \ref{fig:efold} to the average counts (see the text).
Lines show the ratio of models taken by phase-resolved spectral analyses 
in table \ref{tbl:spec_qui} to the average models;
MEKAL + MEKAL, MEKAL + PL, and CEMEKL (Also see the text).}
\label{fig:efold_amplitude}
\end{figure}

First, we tried phase-resolved spectral analyses.
From the spin profile in Fig.\ \ref{fig:efold}, we defined 
the top and bottom phases as epochs at spin phases of 
$\phi$ = 0.59375 -- 1.21875 and 0.21875 -- 0.59375, respectively,
and selected XIS events in these two phases.
The X-ray spectra in both phases are shown 
in Fig.\ \ref{fig:spec_qui_resolved}. 
Following the phase averaged studies in section \ref{section:ana_qui_spec}, 
we tried to fit these spectra with (0) MEKAL, (1) MEKAL+MEKAL, 
(2) MEKAL+PL, and (3) CEMEKL models, 
and got results shown in table \ref{tbl:spec_qui} and 
Fig.\ \ref{fig:spec_qui_resolved}.
As a result, there were no strict differences statistically
among these models except for case-0, single MEKAL model,
as already shown in the phase averaged analyses.
In addition, the X-ray spectra of the pulsation component
(i.e., subtraction from the spectra during the top phase of those
during the bottom phase) were also well represented both by
thermal and non-thermal models as shown in table \ref{tbl:spec_qui_pulse};
i.e., we could not distinguish these origins statistically,
only from the phase-resolved analyses.

Next, we evaluated numerically the amplitudes of spin modulations 
as a function of the energy. 
Since the modulations are well reproduced by a
single sine function (reduced $\chi^2$/d.o.f $\sim 0.6$)
and power of higher harmonics of the waveform are negligible 
in estimating amplitudes, 
we fitted the energy-resolved pulse profiles in Fig.\ \ref{fig:efold}
by a sine function with a constant model, 
and derived the ratios of the pulsation amplitudes to 
the average ones (i.e., a DC component).
As shown in the results in Fig.\ \ref{fig:efold_amplitude} crosses,
it was clearly seen that there were shallower modulations 
in higher energy band.
This situation is different from those seen with {\it ASCA}
in the high state \citep{ishida97}.
Then, in order to check the consistency of this result with 
phase-resolved spectral models,
we also plotted in the same figure the ratios of models
between the pulsation components (i.e. subtractions of 
best fit models in the bottom phase from those in the top) 
and the best fit models in the average (section \ref{section:ana_qui_spec}).
Therefore, MEKAL + MEKAL model (case-1) requires 
a large modulation in the hard X-ray band, 
which was inconsistent with the manner of the spin modulation.
The best model was MEKAL + PL (case-2) in Fig.\ \ref{fig:efold_amplitude}, 
although CEMEKL model (case-3) could not be excluded.
Note that the negative values of the ratio seen in the MEKAL+PL model 
is due to the difference of the best-fit values of the PL index and 
it is consistent with being 0 above 20 keV.

\section{Discussions}
\label{section:discussion}
\subsection{Plasma Parameters during the Flare}
\label{section:discussion_param_flare}
A flare event was observed during the X-ray observation of AM Her 
in the very low state with {\it Suzaku},
as shown in section \ref{section:ana_flare}.
Same kinds of flares were reported in other polars,
such as UZ For during a low X-ray luminosity \cite{pandel02}.
The X-ray flare of AM Her decayed with a time scale of
$\tau_{\rm flare} \sim$ 3700 sec, 
and the X-ray spectra was well reproduced by 
the thermal plasma model with the average temperature 
of $kT = 8.7$ keV (table \ref{tbl:spec_flare}) and 
the volume emission measure (VEM) of 
$3.12\times 10^{53} \left(\frac{d}{91 {\rm pc}}\right)^{2}$ cm$^{-3}$.
In this section, we discuss on a rough estimation of plasma parameters 
during the flare, i.e., the electron density, $n_{\rm e}$, 
and the plasma size (or the shock height of the plasma, $h$),
by using three observational values of $\tau_{\rm flare}, kT,$ and VEM.

First, from the observational parameter of $\tau_{\rm flare}$,
we can estimate the value of $h$, 
if we assume the flare time scale ($\tau_{\rm flare}$) 
is mainly determined by the cooling time scale of the plasma, 
$\tau_{\rm cool}$.
This assumption may be correct because 
the temperature seemed to be gradually decreased 
through the flare as seen in the bottom panel of Fig.\ \ref{fig:flare_lc}.
According to Aizu (1973), the shock height is numerically 
described as $h = 0.605 \cdot u_{\rm s} \cdot \tau_{\rm cool}$,
where $u_{\rm s}$ is a shock velocity of the plasma,
which is about $10^9$ cm s$^{-1}$ 
on a typical gravitational potential of a WD in a strong shock case,
as 
\begin{eqnarray}
  u_{\rm s} =&& 1.2 \times 10^{9} \nonumber \\
     &&\times
          \left(\frac{M_{\rm WD}}{0.88 M_\odot}\right)^{1/2} 
          \left(\frac{R_{\rm WD}}{10^9 {\rm ~cm}}\right)^{-1/2} 
          {\rm cm~s}^{-1}.
\label{eq:u_shock}
\end{eqnarray}
Thus, we get 
\begin{eqnarray}
  h =&& 3.1 \times 10^{11} 
          \left(\frac{\tau_{\rm cool}}{3700{\rm ~s}}\right)\nonumber \\
     &&\times
          \left(\frac{M_{\rm WD}}{0.88 M_\odot}\right)^{1/2} 
          \left(\frac{R_{\rm WD}}{10^9 {\rm ~cm}}\right)^{-1/2} 
          {\rm cm}.
\label{eq:flare_h}
\end{eqnarray}
This value is quite large in the binary system of AM Her
compared to the binary distance, $a$, 
which is calculated by the third law of Kepler as
\begin{equation}
 a = 9.8 \times 10^{10}
         \left(\frac{M_{\rm WD}}{0.88 M_\odot} 
         + \frac{M_2}{0.3 M_\odot}\right)^{1/3}
         \left(\frac{P_{\rm orb}}{P_{\rm XIS}}\right)^{2/3}
         {\rm cm},
\label{eq:flare_a}
\end{equation}
where $M_2$ the mass of the companion star \citep{bailey88}
and $P_{\rm orb}$ is the orbital period of the binary.
In other words, this solution requires that 
the binary space is fully filled with the plasma, and it is unphysical.
Therefore, the cooling time scale should be shorter than 
the flare time scale (i.e., $\tau_{\rm cool} \ll \tau_{\rm flare}$)
and $\tau_{\rm flare}$ is just described 
by a time scale of the accretion onto the WD from the companion star.
This statement is supported by the observational fact 
seen in Fig.\ \ref{fig:flare_lc} top; 
the X-ray flux became faint during the epoch corresponding 
to the dim phase in quiescence, and thus the flare phenomenon
occurs near the magnetic pole on the WD surface.

\begin{figure}[htb]
\centerline{\includegraphics[angle=0,scale=0.30]{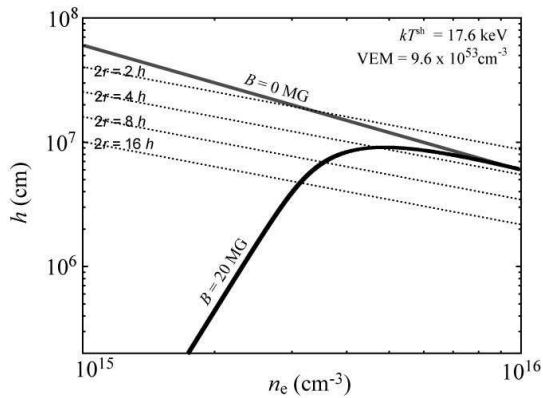}}
\caption{Thick lines show numerical solutions of $h$ and $n_{\rm e}$,
based on calculations on $kT_{\rm sh}$ with {\it Suzaku}
during the flare phase of AM Her in the very low state.
Dashed lines show the limitations from VEM with assumption of
$a = 2, 4, 8, 16$. See the text.}
\label{fig:flare_ne_h}
\end{figure}

Second, from the observational parameter of $kT$,
we can constrain values of $n_{\rm e}$ and $h$,
under the condition that $\tau_{\rm cool}$ is not limited 
by $\tau_{\rm flare}$ as discussed above.
Here, we use $n_{\rm e}$ as the electron density 
just below the shock front, and
we assume that the maximum temperature obtained with
the CEMEKL model, $kT_{\rm max}$, represents the shock 
temperature, $kT_{\rm sh}$, of the multi-color accretion column;
i.e., we assume $kT_{\rm sh} = kT_{\rm max}$.
According to \citet{lamb79},
the cooling of the plasma occurs via free-free emission and 
cyclotron radiation, and the time scale of the cooling is described as
\begin{equation}
 \tau_{\rm cool} = \frac{3 n_{\rm e} kT}
                {2 \sqrt{\bar{\epsilon_{\rm ff}}^2 + \bar{\epsilon_{\rm cyc}}^2}},
 \label{eq:flare_cool}
\end{equation}
where $\bar{\epsilon_{\rm ff}}$ and $\bar{\epsilon_{\rm cyc}}$ are
the volume-emissivities averaged along the accretion column
of the free-free and cyclotron emissions, respectively.
The averaged volume-emissivity of free-free emission are 
quantitatively described as
\begin{eqnarray}
\bar{\epsilon_{\rm ff}} =&& A \times 7.3 \times 10^8 
          \left(\frac{kT_{\rm sh}}{17.6 {\rm ~keV}}\right)^{1/2} \nonumber\\
        &&\times
          \left(\frac{n_{\rm e}}{5 \times 10^{15}{\rm ~cm}^{-3}}\right)^{2}
        \nonumber\\
        && {\rm erg~ sec}^{-1} \rm{~cm}^{-3},
\label{eq:flare_e_ff}
\end{eqnarray}
where $A$ is a correction factor of the plasma structure;
when we put $z$ as a position from the bottom of the plasma,
and the plasma has a vertical dependence of 
$kT \propto n_e^{-1} \propto (z/h)^p$,
then $A$ is described as $\frac{2}{2-3p}$.
The cyclotron emissivity $\bar{\epsilon_{\rm cyc}}$ is given 
in equation (10) of \citet{wu94} 
which depends on $h^{0.85}$.
Here, the value $h$ was determined by $\tau_{\rm cool}$ in 
equation (\ref{eq:flare_cool}),
which has a dependency with $\bar{\epsilon_{\rm cyc}}$ again.
Thus, we need to solve the simultaneous equations 
between  $\bar{\epsilon_{\rm ff}}$,  $\bar{\epsilon_{\rm cyc}}$, 
$h$, and $n_{\rm e}$ to get self consistent values.
We performed numerical calculations to solve them
to derive a rough estimation between $n_{\rm e}$ and $h$.
In the calculation, we assume the Aizu solution (1973)
where the cyclotron cooling was ignored;
i.e., we assume the relation $h \sim 0.6 u_{\rm s} \tau_{\rm cool}$
and set the plasma structure at $p = 2/5$.
We also ignored the effects of the gradient of the gravitational 
potential of the WD in the accretion plasma.
The solutions, 
when the magnetic field strength is a typical value of 
AM Her $B = 20$ MG \citep{schmidt81},
are shown in Fig.\ \ref{fig:flare_ne_h}.

Finally, we can use the third observational value, VEM,
to limit the parameters $h$ and $n_{\rm e}$.
To evaluate the volume of the plasma, 
we assume that the accretion column has a cylinder shape 
with a radius of $r$, which is $a$ times size of the shock height 
(i.e., $2r = a h$).
Dashed lines in Fig.\ \ref{fig:flare_ne_h} represent the limitation
from VEM values to $h$ and $n_{\rm e}$ in cases of $a = 2, 4, 8$, and 16.
In the plot, we used the VEM value from a simple MEKAL fitting;
since the model assumes uniform density and the plot is shown 
in the shock density, we multiplied a correction factor of 
$1/(1-2p)$, which is 4 when we assume the Aizu solution (1973).
As a result, we can get a rough values of $h$ and $n_{\rm e}$ 
as a function of $a$; for example, 
$n_{\rm e} \sim 5 \times 10^{15}$ cm$^{-3}$ and 
$h \sim 10^7$ cm when $a = 4$.
Although many solutions are allowed in this estimation, 
quite small values of $a$ are not realistic 
from Fig.\ \ref{fig:flare_ne_h};
i.e., the plasma has a coin-like shape rather than 
tall-cylinder one.

\subsection{Structure of the Thermal Plasma in the Quiescence}
\label{section:discussion_thermal}

According to the results of the data analyses 
in sections \ref{section:ana_qui_spec} 
and \ref{section:ana_qui_phaseresolved},
the X-ray spectra of AM Her during the quiescence can be described 
either by a multi-color-plasma CEMEKL model with a temperature of 
$kT_{\rm max} = 6.7$ keV, 
or by a combination of thermal and non-thermal model of
MEKAL with $kT = 1.1$ keV plus the PL component.
In this section, we focus on the nature of the thermal component,
whereas non-thermal phenomena in AM Her will be discussed in
the next section \ref{section:discussion_acceleration}.

First, we estimated the plasma parameters, $n_{\rm e}$ and $h$,
during the quiescence by performing numerical calculations 
as already shown in section \ref{section:discussion_param_flare}.
The VEMs during the quiescence were one or two order-of-magnitude 
lower than that in the flare; VEM = $2.5\times 10^{52}$ cm$^3$ (CEMEKL) 
or $1.4\times 10^{51}$ cm$^3$ (MEKAL+PL).
To obtain the value of $kT_{\rm sh}$, 
we assumed $kT_{\rm sh} = kT_{\rm max}$ for CEMEKL model
as in section \ref{section:discussion_param_flare}.
We cannot estimate $kT_{\rm sh}$ directly from $kT$ of MEKAL model,
so we simulated PHA spectra from CEMEKL model 
with various $kT_{\rm max}$ on XSPEC and 
fitted with MEKAL models to obtain a rough relation between 
$kT$ and  $kT_{\rm max}$.
In the simulation, we used the XIS response on the HXD nominal position
and fitted spectra in the hard X-ray band of the XIS, 4 -- 10 keV. 
Then, we got a $kT_{\rm max}$ -- $kT$ relation as 
\begin{equation}
\log(kT_{\rm max}) \sim 1.23 \log(kT) + 0.14.
\label{eq:kt_cemekl_mekal}
\end{equation}
Thus, the value $kT = 1.1$ keV by MEKAL analysis corresponds 
to $kT_{\rm sh} \sim 1.6$ keV of the multi-color column.
Using these VEM and $kT_{\rm sh}$ values, 
we performed the same numerical calculations again 
as in section \ref{section:discussion_param_flare},
and got plasma parameters as 
$n_{\rm e} \sim$ a few $\times 10^{14}$ cm$^{-3}$ and 
$h \sim 10^7$ cm when $a = 4$.
Although there remains an uncertainty on the plasma shape $a$,
the column height seems to be almost the same order of magnitude 
as that in flares,
and the electron density got one or two order of magnitude lower 
than in flares.

\begin{figure}[htb]
\centerline{\includegraphics[angle=0,scale=0.35]{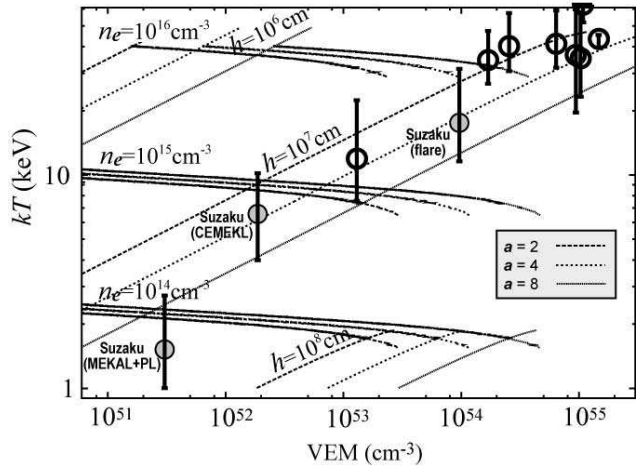}}
\caption{Crosses with circle show scatter plot of $kT$ and VEM, 
obtained by X-ray observations of AM Her
\cite{beardmore95,ishida97,matt00,terada04,girish07}.
Lines show relations between $kT$ and VEM by numerical calculations 
on the various cases of 
$h =10^{6}, 10^{7}, 10^{8}$ cm,
$n_{\rm e} = 10^{14}, 10^{15}, 10^{16}$ cm$^{-3}$,
and $a = 2, 4, 8$. Conditions are shown in the figure.}
\label{fig:qui_kt_em}
\end{figure}

Second, in order to compare the {\it Suzaku} results 
with other past observations in the high, medium, and low states,
we plotted $kT$ and VEM values by several X-ray satellites
\citep{beardmore95, ishida97, matt00, terada04, girish07}
in Fig.\ \ref{fig:qui_kt_em}.
In this figure, the $kT$ values obtained with MEKAL models
are corrected into $kT_{\rm sh}$ equivalents by 
the $kT$-$kT_{\rm max}$ relation of equation (\ref{eq:kt_cemekl_mekal}).
Therefore, observationally, the plasma temperature gets 
lower as VEM gets smaller, and the {\it Suzaku} points 
in the faintest state also follow this trend.
Hereafter, we call this relation as the VEM - $kT$ trend.

Finally, to interpret the VEM - $kT$ trend in Fig.\ \ref{fig:qui_kt_em}
quantitatively, we performed further numerical calculations 
demonstrated in section \ref{section:discussion_param_flare}.
In Fig.\ \ref{fig:qui_kt_em}, we plotted contours of VEM and $kT$ 
with various conditions of $a$, $n_{\rm e}$, and $h$. 
The calculations includes the free-free and cyclotron cooling processes
with several assumptions described 
in section \ref{section:discussion_param_flare}.
Therefore, the results of calculations suggest that 
the changes on the plasma parameters occur only on $n_{\rm e}$, 
keeping the column shape almost constant,
even when the VEM varies by two order of magnitudes.
As shown in Fig.\ \ref{fig:qui_kt_em},
one realistic set of parameters is $a = 4$ and $h = 10^7$ cm, 
which were already presented as an one possible solution 
of the flare plasma in section \ref{section:discussion_param_flare}.
Other possibilities are larger $a$ at lower $h=10^6$ cm
or smaller $a$ at higher $h=10^8$ cm,
but the latter case is less feasible because analytic solutions
vanish at small $a$ values, as already mentioned 
in section \ref{section:discussion_param_flare}.
In other words, our calculation does not exclude other interpretations
that both $n_{\rm e}$ and the plasma shapes varies as VEM changes.

If the plasma shape is constant in various states
keeping the same shock height $h$, 
then the gravitational potentials at the shock front 
are not changed by its accretion rate.
The variation of $kT$ as observed in Fig.\ \ref{fig:qui_kt_em}
may be caused by possible changes of the energy-conversion efficiency
$\eta$ at the shock, however, the adiabatic index of the accretion gas 
is required to be less than 1 to change $\eta$ by factor two or three;
this situation is not realistic.
Another possibility is an observational effect;
X-ray observations detect the free-free emissions and thus 
are essentially sensitive to the dense bottom-part of the column
where the plasma is already cooled down mainly by the cyclotron radiation.
At lower accretion rate, the effect will be significant 
because the cyclotron cooling becomes dominant in the lower densities
as discussed in section \ref{section:discussion_param_flare}.

\subsection{Particle Acceleration on the Accretion Shocks}
\label{section:discussion_acceleration}
In this section, 
we assume that the hard X-ray emission detected with {\it Suzaku} 
has non-thermal origin and 
we concentrate on the particle acceleration processes in AM Her.
As demonstrated in sections \ref{section:ana_qui_spec} and 
\ref{section:ana_qui_phaseresolved}, 
the X-ray luminosity of the power-law component of AM Her was 
$L_{\rm x, PL} = 1.5 \times 10^{29} {\rm ~erg ~sec}^{-1}$ 
at $d=91$ pc in 0.5 -- 10 keV band,
which is the same order of magnitude 
as that of the first WD-pulsar candidate, AE Aqr, at 
$6.6 \times 10^{29} {\rm ~erg ~sec}^{-1}$ at $d=102$ pc \citep{terada08b}. 
Thus, neither bremsstrahlung nor curvature radiations are 
feasible as a physical process of non-thermal emissions
from a viewpoint of radiation efficiencies,
although another possibility is the synchrotron radiation
from about MeV electrons under the strong magnetic field 
of the WD \citep{terada08b}.

If particles are accelerated by an electric potential
induced by the spin rotation of the magnetized WD up to $10^{14}$ volts,
the acceleration process should occur on the same place 
as the non-thermal emissions; 
i.e., near the WD surface where the magnetic field is 
strong enough to create hard X-rays via synchrotron process.
However, in such a polar-cap case, the emissions should 
present spin pulsations like AE Aqr case, 
whereas no modulations were observed in the hard X-ray band
from AM Her (section \ref{section:ana_lc}).
Thus, electric acceleration mechanism may be less feasible,
although we need more theoretical studies.
In this section, we further consider another possibility
of diffusive shock-acceleration mechanism \citep{bell78, blandford78} 
on the accretion shock on the pole.

Diffusive shock acceleration mechanism occurs 
on the collisionless-shock case,
which will be realized when the kinetic energy of the plasma 
particles exceeds the Coulomb potential of the plasma.
The ratio between them are called as the binding parameter, $\zeta$,
and the collisionless shock occurs on the condition of $\zeta < 1$.
The parameter can be described by
\begin{equation}
\zeta \sim 0.18 \left(\frac{n_{\rm e}}{10^{15} {\rm cm}^{-3}}\right)^{1/3} 
\left(\frac{kT_{\rm pre}}{1 {\rm eV}}\right)^{-1},
\end{equation}
where $kT_{\rm pre}$ is a temperature of the pre-shock matters.
Therefore, the accretion shock on the WD pole has a condition
of a collisionless shock. 
In the post-shock plasma, 
according to equation (5.31) by \citet{spitzer62}
the collisional timescale between the 
electrons and ions, $\tau_{\rm ie}$, is given as
\begin{equation}
\tau_{\rm ie} \sim 0.01 \left(\frac{kT_{\rm sh}}{17.6 {\rm ~keV}}\right)^{3/2}
\left(\frac{n_{\rm e}}{5 \times 10^{15}{\rm ~cm}^{-1}}\right)^{-1}\ {\rm s},
\end{equation}
which is one order of magnitude smaller than 
the cooling time scale (typically $\sim$ 0.3 sec)
by equation (\ref{eq:flare_cool}). 
Therefore, the collisionless region exists at the top of
the accretion column, just beneath the shock,
although electrons themselves or protons themselves 
may be in the collisional equilibrium under high densities.
This idea is supported by the fact that 
the shallower spin modulation was observed 
in the harder energy band (section \ref{section:ana_lc}).

We can assume that
the magnetic field is parallel to the shock normal.
In this case,
the acceleration time-scale on the shock ($t_{acc}$) is derived as
\begin{equation}
t_{\rm acc} = \frac{20}{3}\xi\frac{E}{eBu_s{}^2}c\ \ ,
\end{equation}
where $\xi$, $E$, $e$, $B$ are
fluctuation of the magnetic field ($> 1$),
the maximum energy of accelerated electrons,
electric charge, and the magnetic field strength, respectively
\citep{jokipii1987,bamba2003,yamazaki2006}.
Assuming $\xi = 1$ (Bohm limit), $B = 20~{\rm MG}$,
and $u_{\rm s}$ at equation (\ref{eq:u_shock}), 
the $t_{\rm acc}$ becomes
\begin{eqnarray}
t_{\rm acc} && = 2.3\times 10^{-5}
\left(\frac{\xi}{1}\right)
\left(\frac{E}{1~{\rm TeV}}\right)  
\left(\frac{B}{20~{\rm MG}}\right)^{-1} \nonumber\\
&& \times
\left(\frac{u_{\rm s}}{1.2\times10^9~{\rm cm~s^{-1}}}\right)^{-2} {\rm s}\ \ ,
\end{eqnarray}
which is very small compared with
acceleration time scale in young supernova remnants
\citep[e.g.,][]{bamba2005}.
In the actual case of the accretion flow onto the WD,
the magnetic fluctuation may be $\xi \gg 1$, but
if it satisfies $\xi < 10^4$, 
the time scale of the acceleration $t_{\rm acc}$ is 
much shorter than the time scale of the accretion 
$t_{\rm cool}\sim 0.3$ s.

The synchrotron loss time scale ($t_{\rm sync}$) is, on the other hand,
\begin{equation}
t_{\rm sync} = 9.8\times 10^{-13}
\left(\frac{E}{1~{\rm TeV}}\right)^{-1}
\left(\frac{B}{20~{\rm MG}}\right)^{-2} \ \ {\rm s},
\end{equation}
from \citet{rybicki1979}.
The maximum energy of accelerated electron is achieved
when the acceleration time-scale becomes
same to the synchrotron loss time-scale.
The maximum energy of the accelerated electrons is, thus,
\begin{eqnarray}
\left(\frac{E}{1~{\rm TeV}}\right) =&&
2.1\times 10^{-4}
\left(\frac{\xi}{1}\right)^{-1/2}
\left(\frac{B}{20~{\rm MG}}\right)^{-1/2} \nonumber\\
&& \times
\left(\frac{u_{\rm s}}{1.2\times10^9~{\rm cm~s^{-1}}}\right) \ \ .
\end{eqnarray}
Our result imply that
the accretion shock of CVs can accelerate electrons
up to $\sim$GeV when $\xi = 1$, which is seen 
in the cases of young supernova remnants and pulsar wind nebulae.
In cases of $\xi \gg 1$, which work out when the shock is close 
to the WD surface and the magnetic field dominates the dynamics 
of the flow, the maximum energy will be reduced
as the dependency of $\xi^{-0.5}$; 
i.e., $E$ will be $\sim$MeV at $\xi=1000$.
Note that the maximum energy of protons can be $\sim$ 1800 times 
larger than electrons, $\sim$ GeV to TeV.
The cut-off frequency of synchrotron X-rays ($\nu_{cut}$) is
\begin{eqnarray}
\nu_{cut} &=&
1.0\times 10^{18}
\left(\frac{E}{1~{\rm TeV}}\right)^2
\left(\frac{B}{20~{\rm MG}}\right) \
\left(\frac{\xi}{1}\right)^{-1}  \ {\rm Hz} \nonumber \\
&=& 27~{\rm keV}\ \ ,
\end{eqnarray}
which is higher than our observation band,
under an assumption of the Bohm limit condition.

\section{Conclusion}
We have observed the polar, AM Her, in a very low state 
with {\it Suzaku} at the X-ray luminosity of 
$1.7 \times 10^{29} \left(\frac{d}{91 {\rm ~pc}}\right)^2$
erg sec$^{-1}$ in the 0.5 -- 10 keV band.

\begin{enumerate}
\item 
The object shows a flare phenomenon 
reaching the X-ray luminosity of
$6.0 \times 10^{29} \left(\frac{d}{91 {\rm ~pc}}\right)^2$
erg sec$^{-1}$ in the 0.5 -- 10 keV band,
with a time scale of $\sim$ 3700 sec (section \ref{section:ana_lc}).
The X-ray spectra can be reproduced by the thermal-plasma MEKAL model
with the temperature of $8.67_{-1.14}^{+1.31}$ keV 
and the abundance of $0.76_{-0.26}^{+0.28}$ solars
(section \ref{section:ana_flare}).
In discussion on the time scale of the flare, 
we concluded that it should not be the cooling time scale 
of the plasma from constraints on the plasma size and density
(section \ref{section:discussion_param_flare}).

\item 
The X-ray light curves during the quiescence show a clear spin
modulation at 0.1289273(2) days at BJD 2454771.581.
These values are consistent with those in a high state
(section \ref{section:ana_lc}).
The spectra during the quiescence are well represented
by MEKAL + PL model or a single CEMEKL model 
(section \ref{section:ana_qui_spec}).
The phase-resolved analyses also support these two models 
(section \ref{section:ana_qui_phaseresolved}).

\item
From historical X-ray measurements of AM Her in various states,
we found that the temperature is positively correlated with 
the volume emission measure.
The {\it Suzaku} results in a very low state also follow this relation.
We made a rough estimation with a simple numerical calculation,
and found that the $kT$-VEM trend can be interpreted 
by a trajectory with a constant plasma shape 
in various electron densities, as an one of possible solutions
(section \ref{section:discussion_thermal}).

\item
The origin of the non-thermal emission was discussed,
and a shock acceleration process at the shock front 
on the top of the accretion column was checked and
proposed (section \ref{section:discussion_acceleration}).

\end{enumerate}

\section*{Acknowledgement}

The authors would like to thank all the members of 
the {\it Suzaku} team for their contributions 
in the maintenance of instruments and software, 
spacecraft operation, and calibrations.
We thank Dr.\ R.\ Yamazaki from Aoyama Gakuin University
for his useful comments and discussions on the particle acceleration.
We also thank Prof.\ K.\ Makishima from University of Tokyo and
RIKEN for his continuous helps and discussions.
We would like to thank the referees and the editor
for their careful readings and helpful comments in preparing the paper.
This work was supported in part 
by the Grant-in-Aid for Young Scientists (B) 
of the MEXT,No. 19740168 (Y.~T),
and by Grant-in-Aid for Scientific 
Research of the Japanese Ministry of Education, 
Culture, Sports, Science and Technology, No.~22684012 (A.~B.).



\begin{thebibliography}{}
\bibitem[Allen(1973)]{allen73} 
  Allen, C. W., 1973, University of London, Athlone Press, 1973, 3rd ed.
\bibitem[Aizu(1973)]{aizu73} 
  Aizu, K, 1973, Progress of Theoretical Physics, 50, 344
\bibitem[Abada-Simon et al.(1993)]{aeaqr_radio2} 
  {Abada-Simon}, M., {Latchkey}, A., {Bastian}, T.~S., 
  {Bookbinder}, J.~A., \& {Dulk}, G.~A. 1993, \apj, 406, 692
\bibitem[Bamba et al.(2003)]{bamba2003} 
  Bamba, A., Yamazaki, R., Ueno, M., \& Koyama, K.\ 2003, \apj, 589, 827
\bibitem[Bamba et al.(2005)]{bamba2005} 
  Bamba, A., Yamazaki, R.,Yoshida, T., Terasawa, T., \& Koyama, K.\ 
  2005, \apj, 621, 793
\bibitem[Baskill et al.(2005)]{baskill05} 
  Baskill, Darren S.; Wheatley, Peter J.; Osborne, Julian P. 2005, 
  \mnras, 357, 626
\bibitem[Bastian et al.(1988)]{aeaqr_radio1} 
  {Bastian}, T.~S., {Dulk}, G.~A., \& {Chanmugam}, G. 1988, \apj, 330, 518
\bibitem[Bailey et al.(1988)]{bailey88} 
  {Bailey}, J. and {Hough}, J.~H. and {Wickramasinghe}, D.~T., 
  1988, \mnras,233, 395
\bibitem[Beasley et al.(1994)]{mcv_radio2} 
  {Beasley}, A.~J., {Bastian}, T.~S., {Ball}, L., \& {Wu}, K. 
  1994, \aj, 108, 2207
\bibitem[Beardmore et al.(1995)]{beardmore95} 
  Beardmore, A. P., Done, C., Osborne, J. P., Ishida, M., 1995, 
  \mnras, 272, 749
\bibitem[Bhat et al.(1991)]{amher_tev} 
  {Bhat}, C.~L., {Kaul}, R.~K., {Rawat}, H.~S., {Senecha}, V.~K., 
  {Rannot}, R.~C., {Sapru}, M.~L., {Tickoo}, A.~K., \& {Razdan}, H. 
  1991, \apj, 369, 475
\bibitem[Bell(1978)]{bell78} 
  Bell, A.~R., 1978, \mnras, 182, 443
\bibitem[Blandford and Ostriker(1978)]{blandford78} 
  Blandford, R. D. and Stroker, J. P., 1978, \apj, 221, 29
\bibitem[Bond et al.(2002)]{mcv_radio4} 
  {Bond}, H.~E., {White}, R.~L., {Becker}, R.~H., \& {O'Brien}, M.~S. 
  2002,\pasp, 114, 1359
\bibitem[Chanmugam \& Dulk(1982)]{amher_radio1} 
  {Chanmugam}, G. \& {Dulk}, G.~A. 1982, \apjl, 255, L107
\bibitem[Cropper et al.(1998)]{cropper98} 
  Cropper, Mark, Harrop-Allin, M. K., Mason, K. O., Mittaz, J. P. D., 
  Potter, S. B., Ramsay, Gavin,1998, \mnras, 293, 57
\bibitem[de Jager(1994)]{aeaqr_acc}
  {de Jager}, O.~C. 1994, \apjs, 90, 775
\bibitem[de Martino et al.(1998)]{demartino00} 
  de Martino D., G{\"a}nsicke, B.~T., Matt, G. ,Mouchet, M., 
  Belloni, T., Beuermann, K.,Bonnet-Bidaud, J.-M., Mattei, J.,
  Chiappetti, L., and Done, C. 1998, \aa, 333, L31
\bibitem[Done \& Osborn(1997)]{done97} 
  Done, C., \& Osborne, J. P., 1997, \mnras, 288, 649
\bibitem[Dulk et al.(1983)]{amher_radio2} 
  {Dulk}, G.~A., {Bastian}, T.~S., \& {Chanmugam}, G. 1983, \apj, 273, 249
\bibitem[Fukazawa et al.(2009)]{hxdnxb09}
  Fukazawa, Y., Mizuno, Y., Watanabe, S., et al. 2009, \pasj, 61, S17 
\bibitem[G{\"a}nsicke et al.(1995)]{gansicke95} 
  G{\"a}nsicke B. T.,  Beuermann, K., \& de Martino, D.,1995, \aap, 303, 127
\bibitem[Girish et al.(2007)]{girish07} 
  Girish, V., Rana, V. R., Singh, K. P.,2007, \apj, 658, 525
\bibitem[H{\= o}shi(1973)]{hoshi73} 
  H{\= o}shi, R., 1973, Progress of Theoretical Physics, 49, 776
\bibitem[Ishida et al.(1991)]{ishida91} 
  Ishida, M., Silber, A., Bradt, H. V., Remillard, 
  R. A., Makishima, K., Ohashi, T.,1991, \apj, 367, 270
\bibitem[Ishida et al.(1994)]{ishida94}
  Ishida, M., Makishima, K., Mukai, K., and Masai, K., 1994, \mnras, 266, 3671
\bibitem[Ishida et al.(1997)]{ishida97} 
  Ishida, M., Matsuzaki, K., Fujimoto, R., Mukai, K., 
  Osborne, J. P., 1997, \mnras, 287,651
\bibitem[Ishida et al.(2009)]{ishida09} 
  Ishida, M., Okada, S., Hayashi, T., Nakamura, R., 
  Terada, Y., Mukai, K., Hamaguchi, K.,2009, \pasj, 61, S77
\bibitem[Jokipii(1987)]{jokipii1987} 
  Jokipii, J.R.\ 1987, \apj, 313, 842
\bibitem[Kaastra et al.(1996)]{spex} 
  Kaastra, J. S., Mewe, R., Nieuwenhuijzen, H. 1996, 
  the 11th Colloquium on UV and X-ray Spectroscopy of Astrophysical 
  and Laboratory Plasmas ,411
\bibitem[Kafka et al.(2005)]{kafka05} 
  Kafka S., Honeycutt, R., K., Howell, S., B., Harrison, T.,E., 
  2005, \apj, 130, 2852
\bibitem[Kawka et al.(2007)]{kawka07} 
  Kawka, A., Vennes, S., Schmidt, G. D., Wickramasinghe, D. T., 
  Koch, R.,2007, \apj, 654, 499
\bibitem[Koyama et al.(2007)]{xis2007} 
  {Koyama}, K. {et al.} 2007, \pasj, 59, S23
\bibitem[Nakajima et al.(2008)]{xis_sci08} 
  {Nakajima}, H., {Yamaguchi}, H., {Matsumoto}, H.,et al. 2008, \pasj, 60, S1
\bibitem[Lamb \& Masters(1979)]{lamb79} 
  Lamb, D. Q., \& Masters, A. R., 1979, \apj, 234, 117
\bibitem[Liedahl et al.(1995)]{liedahl95} 
  Liedahl, Duane A., Osterheld, Albert L., Goldstein, 
  William H., 1995, \apj, 438, 115
\bibitem[Matt et al.(2000)]{matt00} 
  Matt, G., de Martino, D., G{\"a}nsicke, B. T., Negueruela, I., 
  Silvotti, R., Bonnet-Bidaud, J. M., Mouchet, M., Mukai, K., 2000, 
  \aap, 358, 177
\bibitem[Mason \& Gray(2007)]{mcv_radio5} 
  {Mason}, P.~A. \& {Gray}, C.~L. 2007, \apj, 660, 662
\bibitem[Mewe et al.(1985)]{mewe85} 
  Mewe, R., Gronenschild, E. H. B. M., van den Oord, G. H. J., 
  1985, \aaps, 62, 197
\bibitem[Meintjes et al.(1994)]{aeaqr_tev2} 
  {Meintjes}, P.~J., {de Jager}, O.~C., {Raubenheimer}, B.~C., 
  {Nel}, H.~I., {North}, A.~R., {Buckley}, D.~A.~H., \& {Koen}, C. 
  1994, \apj, 434, 292
\bibitem[Meintjes et al.(1992)]{aeaqr_tev1} 
  {Meintjes}, P.~J., {Raubenheimer}, B.~C., {de Jager}, O.~C., 
  {Brink}, C., {Nel}, H.~I., {North}, A.~R., {van Urk}, G., \& 
  {Visser}, B. 1992, \apj, 401, 325
\bibitem[Mitsuda et al.(2007)]{suzaku07} 
  Mitsuda, K et al., 2007, \pasj, 59, S1
\bibitem[Mukai and Charles(1987)]{mukai87} 
  Mukai, K., and Charles, P. A., 1987, \mnras, 226, 209
\bibitem[Nelson \& Spencer(1988)]{mcv_radio1} 
  {Nelson}, R.~F. \& {Spencer}, R.~E. 1988, \mnras, 234, 1105
\bibitem[Osborne et al.(1986)]{osborne86} 
  Osborne, J. P., Bonnet-Bidaud, J.-M., Bowyer, S., Charles, P. A., 
  Chiappetti, L., Clarke, J. T., Henry, J. P., Hill, G. J., Kahn, S., 
  Maraschi, L., Mukai, K., Treves, A., Vrtilek, S. 1986, \mnras, 221, 823
\bibitem[Pandel et al.(2002)]{pandel02} 
  Pandel, Dirk, C{\'o}rdova, France A. 2002, \mnras, 336, 1049
\bibitem[Patterson(1994)]{mcv_review} 
  Patterson J., 1994, \pasp, 106, 209
\bibitem[Pavelin et al.(1994)]{mcv_radio3} 
  {Pavelin}, P.~E., {Spencer}, R.~E., \& {Davis}, R.~J. 1994, \mnras, 269, 779
\bibitem[Rybicki \& Lightman(1979)]{rybicki1979} 
  Rybicki, G. B., \& Lightman, A. P. 1979, 
  Radiative Processes in Astrophysics (New York: Wiley Interscience)
\bibitem[Rothschild et~al(1981)]{rothschild81} 
  Rothschild, R. E., Gruber, D. E., Knight, F. K., Matteson, J. L., 
  Nolan, P. L., Swank, J. H., Holt, S. S., Serlemitsos, P. J., 
  Mason, K. O., Tuohy, I. R.,1981, \apj, 250, 723
\bibitem[Schmidt et al.(1981)]{schmidt81} 
  Schmidt, G. D., Stockman, H. S., \& Margon, B.,1981, \apj 243, 157
\bibitem[Schmidt et al.(1983)]{schmidt83} 
  Schmidt, G. D., Stockman, H. S., Grandi, S. A.,1983, \apj, 271, 735
\bibitem[Schmidt et al.(2003)]{schmidt03} 
  Schmidt, Gary D.,  Harris, Hugh C.,  Liebert, James,  et~al, 
  2003, \apj, 595, 1101
\bibitem[Spitzer(1962)]{spitzer62}
Spitzer,L., 1962, Physics of Fully Ionized Gases, 
New York Interscience (2nd edition)
\bibitem[Stockman et al.(1977)]{stockman77} 
  Stockman, H. S., Schmidt, G. D., Angel, J. R. P., Liebert, J., 
  Tapia, S., Beaver, E. A., 1977, \apj, 217,815
\bibitem[Tanaka et al.(1994)]{asca1994} 
  Tanaka, Y., Inoue, H. , Holt, S.~S., 1994, \pasj, 46, L37
\bibitem[Takahashi et al.(2007)]{hxd2007a} 
  Takahashi, T., et al.\ 2007, \pasj, 59, S35
\bibitem[Terada et al.(1999)]{terada99} 
  Terada, Y., Kaneda, H., Makishima, K., Ishida, M., et al, 1999, \pasj, 51, 39
\bibitem[Terada et al.(2001)]{terada01} 
  Terada, Y., Ishida, M., Makishima, K., Imanari, T., Fujimoto, R., 
  Matsuzaki, K., Kaneda, H., 2001, \mnras, 328, 112
\bibitem[Terada et al.(2004)]{terada04} 
  Terada, Y., Ishida, M., Makishima, K., 2004, \pasj, 56, 533
\bibitem[Terada et al.(2005)]{terada05} 
  Terada, Y., Watanabe, S., Ohno, M., Suzuki, M. , Itoh, T. , 
  Takahashi, I. , Sato, G. , Murashima, M., Kawano, N., Uchiyama, Y., 
  Kubo, S., Takahashi, T., Tashiro, M., Kokubun, M., Makishima, K., 
  Kamae, T., Murakami, T., Nomachi, M., Fukazawa, Y., Yamaoka, K., 
  Nakazawa, K., and Yonetoku, D., 2005, IEEE TNS 52, 902-909
\bibitem[Terada et al.(2008a)]{terada08a} 
  Terada, Y., Enoto, T., Miyawaki, R., et al.\  2008 \pasj, 60, S25
\bibitem[Terada et al.(2008b)]{terada08b} 
  Terada, Y., Hayashi, T., Ishida, M., Mukai, K., Dotani, T., 
  Okada, S., Nakamura, R., Naik, S., Bamba, A., Makishima, K.,2008 \pasj 60 387
\bibitem[Yamazaki et al.(2006)]{yamazaki2006} 
  Yamazaki, R., Kohri, K., Bamba, A., Yoshida, T., Tsuribe, T., \& 
  Takahara, F.\ 2006, \mnras, 371, 1975
\bibitem[Young et al.(1981)]{young81} 
  Young, P., Schneider, D. P., Shectman, S. A., 1981, \apj, 245, 1043
\bibitem[Wickramasinghe \& Martin(1985)]{wichramasinghe85} 
  Wickramasinghe, D. T. \& Martin, B., 1985, \mnras, 212, 353
\bibitem[Wickramasinghe et al.(1984)]{wichramasinghe84} 
  Wickramasinghe, D. T., Visvanathan, N., \& Tuohy, I. R.,1984, \apj, 286, 328
\bibitem[Wu, Chanmugam, and Shaviv(1994)]{wu94} 
  Wu K., Chanmugam, G., Shaviv, G., 1994, \apss, 322, 71
\end{thebibliography}
\end{document}